\documentclass[preprint,3p,times]{elsarticle}

\usepackage[utf8]{inputenc}
\usepackage{CJKutf8}
\usepackage{amssymb}
\usepackage{amsmath,amssymb}
\usepackage{algpseudocode}
\usepackage{algorithm}
\usepackage{url}
\usepackage{graphicx}
\usepackage{comment}
\usepackage{bm}
\newcommand{\vect}[1]{\boldsymbol{\mathbf{#1}}}
\def\vec#1{\vect{#1}}
\usepackage{lineno}
\usepackage{color}
\newcommand\HL[1]{{\color{black}#1}}
\newcommand\HLL[1]{{\color{black}#1}}
\newcommand\HLLL[1]{{\color{black}#1}}

\journal{journal}

\begin{document}

\begin{frontmatter}

\title{
Reproduction of vortex lattices in the simulations of rotating liquid helium-4 by numerically solving the two-fluid model using smoothed-particle hydrodynamics incorporating vortex dynamics}
	
\author[RCAST]{Satori Tsuzuki}
\ead{tsuzukisatori@g.ecc.u-tokyo.ac.jp}

\address[RCAST]{Research Center for Advanced Science and Technology, The University of Tokyo, 4-6-1, Komaba, Meguro-ku, Tokyo 153-8904, Japan}

\begin{abstract}
Our recent study has shown that the representative phenomena of liquid helium-4 rotating in a cylinder could be simulated by solving the two-fluid model using smoothed-particle hydrodynamics (SPH) after reformulating the viscosity to conserve the rotational angular momentum. Specifically, the emergence of multiple parallel vortices and their rigid-body rotations were observed in our previous SPH simulations. The reported scheme is based on a classical approximation that assumes the fluid forces of both components and their interactions, with the expectation of functioning as a coarse-grained model of existing approximations that couple a microscopic model and the Navier--Stokes equation. Based on previous studies, this paper proposes an improved SPH scheme that explicitly incorporates vortex dynamics into SPH to reproduce \HL{vortex} lattices, which was not possible in previous studies. Consequently, our improved scheme was observed to reproduce vortex lattices by introducing the Magnus force and interaction forces among vortices into the reformulated two-fluid model. The spinnings of the vortices and rigid-body rotations were also observed. The number of vortices showed a certain agreement with Feynman's rule after the model parameter was optimized. Notably, from a scientific point of view, such vortex lattices are reproduced by the classical-mechanical approximation. 
We hope that our model will help physicists studying low-temperature physics find a new way of approaching this bizarre phenomenon that has attracted attention for more than 80 years.  
\end{abstract}


\end{frontmatter}







\section{Introduction}
The two-fluid model proposed by Tiza and Landau has been recognized as the phenomenological governing equation for superfluid helium-4~\cite{TISZA1938, PhysRev.60.356}. To date, several physicists have discussed the interesting properties of the two-fluid model. In the mid-\HLLL{20}th century, Gorter and Mellink observed a mutual friction force between the normal and superfluid components of the two-fluid model~\cite{GORTER1949285}. Currently, it is generally accepted to combine the Navier--Stokes equation, which represents the normal fluid component, with the vortex filament model, which represents the superfluid component in a Lagrangian form~\cite{Idowu2001, PhysRevLett.120.155301, doi:10.1063/1.5091567, PhysRevLett.124.155301}. 
This suggests that solving both equations of the two-fluid model simultaneously can also provide an approximation, although such an approximation is not correct from a quantum mechanical perspective because fluid forces are assumed for both components. Nevertheless, even if we attempt to numerically reproduce macroscopic phenomena, such as the ``fountain effect'' and ``film flow effect,'' the scale of the problem is so significant that existing quantum mechanical models cannot be directly applied. In addition, these phenomena do not always follow classical fluid mechanics. Therefore, we need to establish a new numerical model assuming that these large-scale problems exist in the boundary region between classical and quantum mechanics.

Accordingly, we have recently developed a numerical scheme that discretizes both components of the two-fluid model using smoothed-particle hydrodynamics (SPH) and solved these components simultaneously~\cite{Tsuzuki_2021}. Surprisingly, it was observed that several rotation phenomena of liquid helium-4 were reproduced despite a rough approximation of the solution of the problems in the quantum scale. Specifically, it was observed that the emergence of multiple vortices rotating parallel to the rotation axis of the outer cylinder, and the phenomena of rigid-body rotation of these vortices, can be reproduced by solving the two-fluid model after reformulating its viscosity term to conserve the angular momentum of particles around their axes (which is often referred to as ``spin angular momentum'' as an analogy for the corresponding term in quantum mechanics). These findings are interesting because they were previously believed to be reproduced only through quantum mechanical approaches. 
However, the numerical model in~\cite{Tsuzuki_2021} could not reproduce the phenomena of so-called \HLL{vortex} lattices because it does not consider the effect of vortex dynamics. Namely, the Magnus force, which is the lift force acting on the vortices caused by their own rotations, the interaction force among parallel vortices, and the effect of quantization of circulation, were not formulated in our previous SPH model. Therefore, based on previous studies, this study improves our SPH model by incorporating these major effects of quantum fluids, to facilitate the reproduction of \HL{vortex} lattices in numerical simulations. 

To the best of our knowledge, no studies other than our previous study numerically solved the two-fluid model to reproduce the dynamics of liquid helium-4 in low-temperature regions, assuming the fluid forces of both components and their interactions. Additionally, no studies have attempted to incorporate the vortex dynamics of a quantum fluid into the SPH formulation. \HLLL{From an application perspective, this study is a preliminary step towards realizing a direct numerical simulation of a bulk quantum liquid. To date, liquid helium-4 has played a significant role in space science applications, for example, the cryogenic cooling systems of X-ray astronomy satellites~\cite{YOSHIDA201827, EZOE2020103016} or space telescopes~\cite{Woods2020, 10.1117/1.JATIS.7.1.011008}. Real-scale simulations of superfluid or near-superfluid liquid helium inside a large-scale cryogenic cooling system can streamline the development processes of these fluids and cooling systems and realize safer and more secure operations. However, as stated, directly solving the microscopic relationships, such as the Gross-Pitaevskii equations~(a nonlinear $\rm Schr\ddot{o}dinger$ equation for boson particles) or considering the effect of Bose-Einstein condensates for all atoms constituting the bulk fluid, remains unrealistic; therefore, there has been a strong demand for exploring a new methodology based on continuum mechanical approximations that incorporates quantum mechanical effects to predict and control the dynamics of large-scale liquid helium-4.}

The remainder of this paper is organized as follows. In Section~2, we briefly review the SPH discretization of the two-fluid model proposed in our previous study. In Section~3, we describe the vortex dynamics in SPH formalism, and derive the SPH formulations of the three representative effects of quantum fluids. In Section~4, we describe the key computational algorithms that realize vortex detection in SPH simulations. In Section 5, we demonstrate the numerical simulations of rotating liquid helium-4 using our SPH model. In Section~6, we summarize and conclude this paper.

\section{Brief review of SPH formulation of the two-fluid model} \label{seq:briefoverviewsphtwofluidm}
Our previous paper~\cite{Tsuzuki_2021} proposed a reformulation of the viscosity term in the motion equations of the two-fluid model to conserve the angular momentum of particles around their axes. 
This idea was inspired by the research by D. W. Condiff, who derived the Navier--Stokes equations with spin angular momentum conservation to reproduce the small effect of rotating molecules for polar fluids ~\cite{doi:10.1063/1.1711295}. As a related study, Condiff's model was introduced by K. M{\"{u}}ller to the development of a numerical scheme for smoothed dissipative particle dynamics for mesoscale heat flows~\cite{MULLER2015301}. In this section, we briefly overview the SPH model of the two-fluid model with spin angular momentum conservation.

\subsection{Two-fluid model with spin angular momentum conservation}
The two-fluid model obtained after reformulating the viscosity term to conserve the spin angular momentum can be expressed as follows~\cite{Tsuzuki_2021}:
\begin{eqnarray}
\rho_{s} \frac{{\rm D} \vec{v}_{s}}{{\rm D} t} &=& -\frac{\rho_{s}}{\rho}\nabla P + \rho_{s}\sigma\nabla T - \vec{F}_{sn}, \label{eq:goveqsuper:mut}\\
\rho_{n} \frac{{\rm D} \vec{v}_{n}}{{\rm D} t} &=& -\frac{\rho_{n}}{\rho}\nabla P - \rho_{s}\sigma\nabla T + (\bar{\eta} + \bar{\eta_{r}})\nabla^2 \vec{v}_{n}  
		+ \biggl( \frac{\bar{\eta}}{3} + \bar{\xi} -\bar{\eta_{r}} \biggr) \nabla\nabla\cdot\vec{v}_{n} + 2\bar{\eta_{r}}\nabla\times\vec{\omega} + \vec{F}_{sn}. \label{eq:goveqnormal:mut}
\end{eqnarray}
Here, $D\{\cdot\}/Dt$ represents the material derivative. $\rho_{n}$ and $\rho_{s}$ are the mass densities of the normal fluid and superfluid components, respectively, which have the relationship $\rho = \rho_{n} + \rho_{s}$, where $\rho$ is the mass density of the total fluid. $\vec{v}_{n}$ and $\vec{v}_{s}$ are the velocities of the normal fluid and superfluid components, respectively. $T$, $P$, and $\sigma$ are the temperature, pressure, and entropy density, respectively. The parameters $\bar{\eta}$, $\bar{\xi}$, and $\bar{\eta_{r}}$ indicate the shear viscosity, bulk viscosity, and rotational viscosity, respectively, where $\bar{\eta_{r}}$ determines the strength of the rotational forces~\cite{MULLER2015301}. Note that the fourth term on the right-hand side of Eq.~(\ref{eq:goveqnormal:mut}) becomes $\rm 0$ because of the incompressibility condition $\nabla \cdot \vec{v} = 0$. Equation~(\ref{eq:goveqnormal:mut}) can be understood as an extended version of the ordinary two-fluid model: the set of the third to fifth terms converges to $\bar{\eta}\nabla^2 \vec{v}_{n}$ as the parameter $\bar{\eta_{r}}$ converges to $0$ in the case of $\nabla \cdot \vec{v} = 0$. This indicates that $\bar{\eta}$ corresponds to the viscosity $\eta_{n}$ of the normal fluid component of the ordinary two-fluid model. Meanwhile, $\vec{F}_{sn}$, the mutual friction force between the two components, is expressed as $\vec{F}_{sn} = 2/3 \rho_{s} \alpha \kappa L \vec{v}_{sn}$, where $\kappa$ is the quantum of circulation, $\vec{v}_{sn}$ is the relative velocity of $\vec{v}_{s} - \vec{v}_{n}$, and $\alpha$ is the friction coefficient. $L$ is the vortex line density, which is time dependent. 
\HLL{Because the exact state of $L$ in rotating liquid helium-4 is unclear, we use} 
a decay model of $L$ \HLL{in counterflow} as an alternative, which \HLL{is} expressed as $L^{-1}(t) = L^{-1}(0) + \beta_{v} t,$ where $\beta_{v}$ is a coefficient in Vinen's equation~\cite{NEMIROVSKII201385}. 

In addition to Eq.~(\ref{eq:goveqsuper:mut}) and Eq.~(\ref{eq:goveqnormal:mut}), we introduce two auxiliary equations: the relationship between the entropy $\sigma$ and the temperature $T$ derived from thermodynamics and the elementary excitation model~\cite{Adamenko_2008, schmitt2015introduction}, and Tait's equation of state that relates the pressure $P$ with the density $\rho$ for each particle. Each equation is expressed as follows ~\cite{schmitt2015introduction}:
\begin{eqnarray}
\sigma &\simeq& \frac{1}{NM}\Biggl[\frac{2\pi^2 k_{B}^4 T^{3}}{45 \hbar^3 c^{3}} + \biggl(\frac{k_{B}}{2\pi}\biggr)^{3/2}\frac{\sqrt{\mu}p_{0}^2 \Delta}{\hbar^3}\frac{e^{-\Delta/T}}{\sqrt{T}}\Biggr], \label{eq:goldentro} \\
P &=& p_{0}\Biggl(\frac{\rho}{\rho_{0}}\Biggr)^{\alpha_{P}} - \beta_{P}. \label{eq:prtcpres}
\end{eqnarray}
Here, in Eq.~(\ref{eq:goldentro}), $\pi$ is the circumference of the circle, $c$ is the speed of sound, $N$ is the total number of helium-4 atoms in the system, and $M$ is the mass of a helium-4 atom. $k_{B}$ represents the Boltzmann constant, and $\hbar$ indicates the reduced Planck's constant.
The constant values of $\mu$, $p_{0}$, and $\Delta$ are given as ${\rm 1.72\times10^{-24}~g}$, ${\rm 2.1\times10^{-19}~gcms^{-1}}$, and ${\rm 8.9~K}$, respectively ~\cite{schmitt2015introduction, bennemann2013novel}.
Note that Eq.~(\ref{eq:goldentro}) holds when $T \ll {\rm 93~K}$.
Meanwhile, in Eq.~(\ref{eq:prtcpres}), $\rho_0$ and $p_{0}$ represent the density and pressure of the fluid in the initial state, respectively. $\alpha_{P}$ determines the incompressibility of the system. $\beta_{P}$ is the reference pressure, which typically corresponds to $p_{0}$. 
As Eq.~(\ref{eq:goldentro}) represents the relationship between the entropy density and the temperature in a quantum ideal gas, we correct $\sigma$ as $\sigma=C_{e}\sigma_{0}$, where $C_{e}$ is the volume ratio of liquid to gaseous helium-4 under the atmospheric pressure $\rm 1.01325~bar$ and it becomes $1.428\times 10^{-3}$~\cite{hammond2000elements}. 
Under constant temperature conditions, we calculated $\sigma_{0}$ using Eq.~(\ref{eq:goldentro}) and retained its value during the simulation to satisfy entropy conservation. These four equations represent the governing equations of the system.

\subsection{SPH discretization}
The basic concept of SPH is an approximation of the Dirac delta function $\delta$ in the integral form of the physical value $\phi$ by the distribution function $W$ called the smoothing kernel function, which converges to $\delta$ when the kernel radius $h$ approaches $\rm 0$, where $h$ is a parameter that determines the scale of the function~($\lim_{h \rightarrow 0} W = \delta$). 
$W$ must also satisfy the normalization condition~($\int W d\vec{r}=1$) and the symmetry $W(-\vec{r}) = W(\vec{r})$. The Gaussian function is an intuitive example of the selection of $W$. This study uses the third-order spline kernel function~\cite{1985AA149135M} because it satisfies a compact support condition in which the value of $W$ becomes zero at a distance of $2h$. 

The physical value $\phi$ expressed in the integral form using $W$ is further discretized based on the summation approximation concept. Under this approximation, $\phi$ is expressed as follows:
\begin{eqnarray}
\phi(\vec{r}_{i}) &\approx& \sum^{N_{p}}_{j}\frac{\phi(\vec{r}_{j})}{\rho_{j}} m_{j} W(|\vec{r}_{i} - \vec{r}_{j}|, h), \label{eq:densityformula}
\end{eqnarray}
where $m_{j}$ and $\rho_{j}$ are the mass and density of the $j$th particle, respectively. $N_{p}$ represents the number of fluid particles used to discretize the system. The gradient, Laplacian, and rotation models of a physical quantity can be obtained axiomatically from Eq.~(\ref{eq:densityformula}) using a vector analysis. The resulting formulas are expressed as follows~\cite{MULLER2015301, doi:10.1146/annurev.aa.30.090192.002551}:
\begin{eqnarray}
\nabla \phi(\vec{r}_{i}) 
	&:=& \rho_{i}\sum^{N_{p}}_{j} m_{j} 
	\Bigl(\frac{\phi(\vec{r}_{i})}{\rho_{i}^2} + \frac{\phi(\vec{r}_{j})}{\rho_{j}^2}\Bigr) 
			\nabla W_{ij},\label{eq:gradient} \\
\nabla^{2} \phi(\vec{r}_{i}) 
	&:=& \sum^{N_{p}}_{j} \frac{m_{j}}{\rho_{j}}\frac{\phi(\vec{r}_{i})-\phi(\vec{r}_{j})}{|\vec{r}_{i}-\vec{r}_{j}|^{2}}  
			(\vec{r}_{i} - \vec{r}_{j})\cdot \nabla W_{ij},\label{eq:laplacian} \\
(\nabla \times \vec{G})_{i} &:=& \sum^{N_{p}}_{j} \frac{m_{j}}{\rho_{j}}\nabla W_{ij} \times (\vec{G}_{i} + \vec{G}_{j}). \label{eq:rotforce}
\end{eqnarray}
Here, $W_{ij} = W(|\vec{r}_{i} - \vec{r}_{j}|, h)$, and $\vec{G}_{i}$ represents a vector value at the position $\vec{r}_{i}$. In the derivation of Eq.~(\ref{eq:gradient}), we used the relationship $\nabla f = \rho[\nabla(f/\rho)+(f/\rho^2)\nabla \rho]$ to ensure the symmetry of $\phi$ between the $i$th and $j$th particles. The temperature gradient and viscosity terms in Eq.~(\ref{eq:goveqsuper:mut}) and Eq.~(\ref{eq:goveqnormal:mut}) are computed using Eq.~(\ref{eq:gradient}) and Eq.~(\ref{eq:laplacian}). On the other hand, the pressure gradient terms in Eq.~(\ref{eq:goveqsuper:mut}) and Eq.~(\ref{eq:goveqnormal:mut}) are computed using an improved scheme~\cite{HU2006844} instead of using Eq.~(\ref{eq:gradient}) to ensure the continuity of the pressure and pressure gradients at the interface.
The fifth term in Eq.~(\ref{eq:goveqnormal:mut}), which provides a rotational force to the normal fluid components, is computed using Eq.~(\ref{eq:rotforce}). The magnitude of the initial angular velocity $|\vec{\omega}|$ is given by $C_{\omega} |\vec{v}_{max}|/0.5d_{p}$, where $d_{p}$ is the initial particle distance, $\vec{v}_{max}$ is the estimated maximum velocity, and $C_{\omega}$ is a scale parameter for $\vec{v}_{max}$.

The mutual frictional force $\vec{F}_{s}^{(i)}$ acting on the $i$th particle, which is a superfluid component, is expressed as follows by using the two-way coupling technique~\cite{ROBINSON2014121, HE2018548}:
\begin{eqnarray}
\vec{F}_{s}^{(i)} &=& C_{L} \frac{\sum_{j \in \Omega_{i}} (\vec{v}_{s}^{(i)} - \vec{v}_{n}^{(j)}) W_{ij}}{\sum_{j \in \Omega_{i}} W_{ij}}.\label{eq:mutualdisc}
\end{eqnarray}
Here, $C_{L}=2/3\rho_{s} \alpha \kappa L$, and $\Omega_{i}$ is the set of all particles in the neighborhood of the $i$th fluid particle. The third term on the right-hand side of Eq.~(\ref{eq:goveqsuper:mut}) was calculated by multiplying Eq.~(\ref{eq:mutualdisc}) using $\rm -1$.
The mutual frictional force $\vec{F}_{n}^{(j)}$ acting on the $j$th particle, which is a normal fluid component, is calculated as the sum of the reaction forces of the force acting on the $i$th particle, which is a superfluid component, as follows:
\begin{eqnarray}
\vec{F}_{n}^{(j)} &=& - \sum_{i \in \Omega_{j}} \vec{F}_{s}^{(i)} \Bigl( \frac{W_{ij}}{\sum_{k \in \Omega_{i}} W_{ik}} \Bigr), \label{eq:mutualdisc:normal}
\end{eqnarray}
where $\Omega_{j}$ is the set of all particles in the neighborhood of the $j$th fluid particle, and $\Omega_{i}$ is the corresponding set for the $i$th particle.
Equation~(\ref{eq:mutualdisc:normal}) corresponds to the sixth term on the right-hand side of Eq.~(\ref{eq:goveqnormal:mut}). 
The right-hand sides of Eq.~(\ref{eq:goveqsuper:mut}) and Eq.~(\ref{eq:goveqnormal:mut}) are discretized. 
Finally, the material derivatives of the velocities on the left-hand sides of Eq.~(\ref{eq:goveqsuper:mut}) and Eq.~(\ref{eq:goveqnormal:mut}) are calculated using the velocity Verlet algorithm, an explicit time-integrating scheme~\cite{PhysRev.159.98}. 
For more details on the SPH model of the two-fluid model with spin angular momentum conservation, refer to the literature~\cite{Tsuzuki_2021}.

\begin{figure}[t]
\vspace{-2.8cm}
\hspace{2.5cm}
\centerline{\includegraphics[width=1.0\textwidth, clip, bb= 0 0 1280 720]{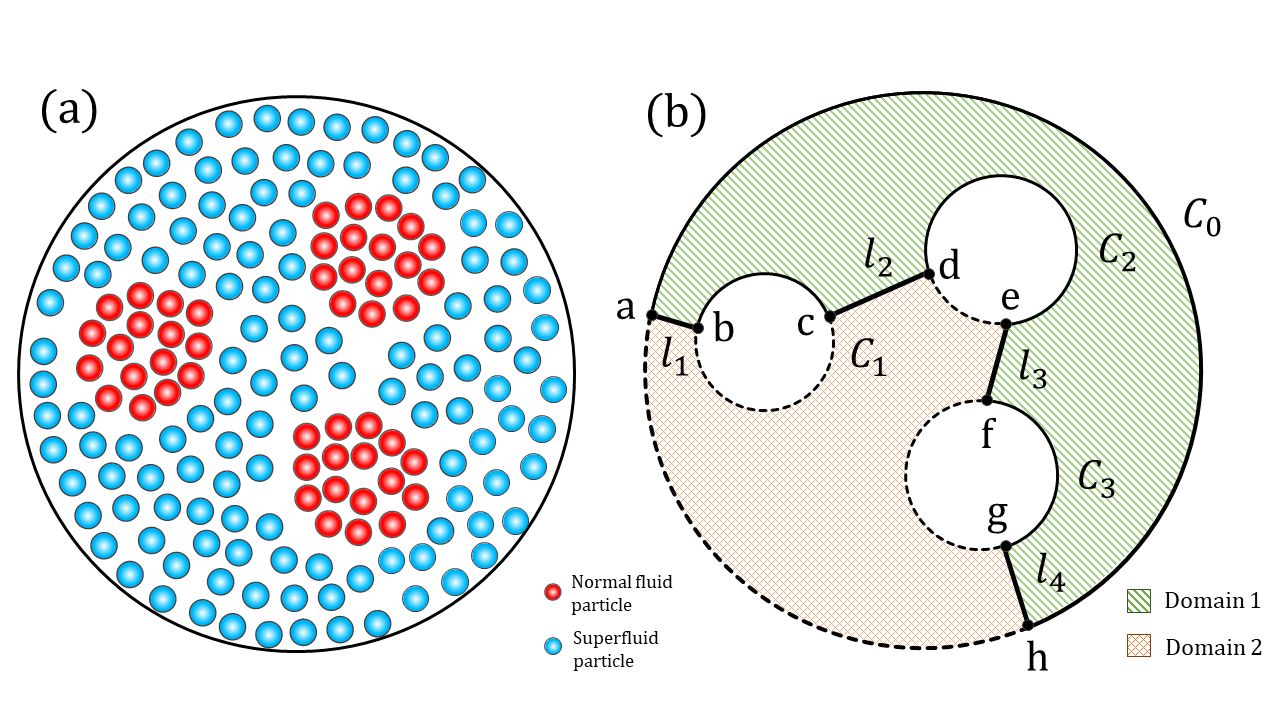}}
\caption{\HLLL{(a)} Schematic diagram showing the geometric correspondence between the characteristic particle arrangement observed in the SPH simulation, and \HLLL{(b)} the Cauchy--Goursat theorem in the complex function theory}
\label{fig:Figure_Cauchy-Goursat}
\end{figure}

\subsection{Numerical techniques for simulation}\label{sec:numericaltechniques}
\HLLL{It is necessary to introduce several established techniques to ensure that the solutions are sufficiently accurate. First, we use a pairwise particle collision method~\cite{SHAKIBAEINIA201213}, which provides a repulsive force when two particles are too close to each other. Second, we introduce a one-dimensional Riemann solver~\cite{ZHANG2017605} for each pairwise particle to avoid using artificial viscosities. These two supportive techniques are effective only when the two particles are closer to each other than the initial particle distance $d_{p}$ as expected. We utilize fixed ghost particles~\cite{COLAGROSSI2003448, MARRONE20111526} to ensure no-slip boundary conditions at the interface between the wall and fluid particles. 
	
In addition, the non-uniformity of the distribution of particles causes particle shortages in the local spaces, which reduces numerical accuracy and causes unphysical pressure oscillations. Therefore, it is important to utilize stabilization techniques. Several filtering techniques have been reported for SPHs with explicit time-integrating schemes. In particular, a density filtering technique known as the Shepard filter for zeroth-order correction among neighboring particles was presented~\cite{pannizo2004physical}. This study adopts a pressure interpolant with a formulation that is similar to the Shepard filter, following the literature~\cite{imoto2019convergence}. Meanwhile, we set the time increment $\delta t$ to be smaller than the three numerical stability conditions: the Courant--Friedrichs--Lewy (CFL) condition and the diffusion and body force conditions~\cite{XU201643}.
To validate the combination of the aforementioned numerical techniques, the driven flows of normal fluid in a square cavity were demonstrated in two cases, $Re$~=~$\rm{100}$ and $Re$~=~$\rm{1,000}$, where $Re$ represents the Reynolds number. For details, refer to~\cite{Tsuzuki_2021}.}
\HLLL{Finally, we implemented our SPH scheme on a single GPU using C/C++ and CUDA~\cite{4541126}. For reference, readers can find an open-source framework recommended as an alternative tool for developing SPH implementation in~\cite{ramachandran2013pysph}. Even in this case, users of the framework still need to explicitly implement the SPH operators and the aforementioned numerical techniques: the pair-wise collision technique, Riemann solver, and pressure interpolant, by themselves.}

\section{Expression of vortex dynamics in SPH formalism} \label{sec:expressionvdsphform}
In the SPH simulations of rotating liquid helium-4 reported in~\cite{Tsuzuki_2021}, the normal fluid particles forming low-density components gather to form clusters and change to vortices, spontaneously rotating around their respective axes. This is illustrated in the schematic diagram in Fig.~\ref{fig:Figure_Cauchy-Goursat}a in an ideal case with three vortices in the simulation domain. Notably, the simulation domain is geometrically divided into two types of domains: the closed domains filled with superfluid particles and the closed pores filled with normal fluid particles. First, we provide a mathematical explanation for the characteristic arrangement shown in Fig.~\ref{fig:Figure_Cauchy-Goursat}a.

\subsection{Description of quantization of circulation in terms of the Cauchy--Goursat theorem}
In the quantum hydrodynamics of liquid helium-4, the velocity of superfluid helium-4 $\vec{v}_{s}$ has the relationship $\vec{v}_{s} = \hbar/m \nabla \theta$, where $\hbar$ is the reduced Planck's constant, $m$ is the mass of a helium-4 atom, and $\theta$ is the phase of the wave function determined by the Gross--Pitaevskii equation. The contour integral of the velocity in a simply connected region ${C}_{s}$ filled only with superfluid components always becomes zero~\cite{pethick2008bose}:
\begin{eqnarray}
\oint_{{C}_{s}} \vec{v}_{s}\cdot d\mathbf{r} &=& 0. \label{eq:quantumcirzero}
\end{eqnarray}
The contour integral of the superfluid velocity in the multiply connected domain $C_{m}$, including the region in which the superfluid component does not exist, corresponds to an integral multiple of the quantum of circulation $\kappa$:
\begin{eqnarray}
\oint_{{C}_{m}} \vec{v}_{s}\cdot d\mathbf{r} &=& q \kappa. \label{eq:quantumcirkappa}
\end{eqnarray}
Here, $q$ represents an integer called the winding number, which mostly becomes $\rm 1$ to stabilize the internal energy of a vortex~\cite{pethick2008bose}.

In the case illustrated in Fig.~\ref{fig:Figure_Cauchy-Goursat}a, a relationship similar to the Cauchy--Goursat theorem holds. We discuss this relationship using Fig.~\ref{fig:Figure_Cauchy-Goursat}b. From Eq.~(\ref{eq:quantumcirzero}), both the contour integrals of the velocity on the boundary of Domain~1 (the route $\tt abcdefgha$ including the solid curves) and that on the boundary of Domain~2 (the route $\tt ahgfedcba$ including the dotted curves) become $\rm 0$. Hence, the following relationship is established:
\begin{eqnarray}
\oint_{\partial D_{1}} \vec{v}_{s}\cdot d\mathbf{r} + \oint_{\partial D_{2}} \vec{v}_{s}\cdot d\mathbf{r} &=& 0. \label{eq:quantumcird1d2A}
\end{eqnarray}
Here, $\partial D_{x}$ represents the boundary of Domain~X, and the direction of the integral path is defined as positive in the counterclockwise direction.
In the integrals on the left-hand side of Eq.~(\ref{eq:quantumcird1d2A}), the line integrals of $l_1$, $l_2$, $l_3$, and $l_4$ are offset between the first and second terms because the integration paths are opposite to each other. 
Consequently, only the line integrals of the velocity on the closed paths surrounding each vortex in the clockwise direction and that on the outermost circle in the counterclockwise direction remain:
\begin{eqnarray}
\sum^{3}_{k=1} \oint_{-\partial C_{k}} \vec{v}_{s}\cdot d\mathbf{r} + \oint_{\partial C_{0}} \vec{v}_{s}\cdot d\mathbf{r} &=& 0. \label{eq:quantumcird1d2B}
\end{eqnarray}
Here, $\partial C_{k}$ represents the closed paths surrounding the $k$th vortex counterclockwise.
Equation~(\ref{eq:quantumcird1d2B}) holds in the general case where the number of vortices is $N_{v}$, which can be expressed as follows:
\begin{eqnarray}
\oint_{\partial C_{0}} \vec{v}_{s}\cdot d\mathbf{r} &=& \sum^{N_{v}}_{k=1} \oint_{\partial C_{k}} \vec{v}_{s}\cdot d\mathbf{r}, \label{eq:quantumcirgeneral}
\end{eqnarray}
where we use the relationship $\oint_{-C} \vec{A}\cdot d\vec{r} = -\oint_{C} \vec{A}\cdot d\vec{r}$. 

Consequently, the circulation of the entire system is given by the sum of the circulations around each vortex.
This result indicates that the case shown in Fig.~\ref{fig:Figure_Cauchy-Goursat}a obtained in our SPH simulations is consistent with the general understanding of multiple parallel quantum vortices.
Furthermore, the fact that the value of the line integral on $C_{i}$ is converted into the area integral of the region surrounded by $C_{i}$ using Stokes' theorem suggests that the total circulation in the area surrounded by $C_{i}$ is distributed to the normal fluid particles constituting the $C_{i}$ vortex. Let us proceed with a discussion building on this mechanical picture. 

\subsection{Circulation in SPH formalism}
Let us define the circulation of the $k$th vortex $\Gamma_{k}$ as follows:
\begin{eqnarray}
\Gamma_{k} &=& \oint_{\partial C_{k}} \vec{v}\cdot d\mathbf{r}, \label{eq:quantumcircond}
\end{eqnarray}
where $\vec{v}$ represents the velocity of the normal or superfluid component. 
If the pore of the $k$th vortex is discretized into $N^{(k)}_{p}$ number of normal fluid particles, Eq.~(\ref{eq:quantumcircond}) is rewritten as follows using Stokes' theorem:
\begin{eqnarray}
 \Gamma_{k} &=& \oint_{\partial C_{k}} \vec{v}\cdot d\mathbf{r}~~=~~\oint_{C_{k}} \nabla \times \vec{v}\cdot d\vec{S} \nonumber \\
			&\simeq& \sum^{N^{(k)}_{p}}_{\alpha = 1} (\nabla \times \vec{v})_{\alpha} \cdot \vec{n}_{\alpha} dS_{\alpha}, \label{eq:quantumcirdiscstokes}
\end{eqnarray}
where $N^{(k)}_{p}$ is the number of particles constituting the $k$th vortex, and $d\vec{S}$ is the surface element vector. $dS_{\alpha}$ is the absolute value of the discretized $d\vec{S}$ defined on the $\alpha$th particle in the $k$th vortex. $\vec{n}_{\alpha}$ is the normal vector of $d\vec{S}_{\alpha}$. We used the relationship $\vec{n}_{\alpha}dS_{\alpha}=d\vec{S}_{\alpha}$ in the derivation of Eq.~(\ref{eq:quantumcirdiscstokes}). 

In SPH, all the particles are equal in size and do not deform. Hence, the particles can be approximated as rigid bodies. In this case, the value of $(\nabla \times \vec{v})_{\alpha}$ is approximately $2\vec{\omega}_{\alpha}$, where $\vec{\omega}_{\alpha}$ is the angular velocity of the $\alpha$th particle. 
Furthermore, in the case of two-dimensional simulations, the value of $dS_{\alpha}$ is equivalent to the small volume of the $\alpha$th particle ($dS_{\alpha} = m_{\alpha}/\rho_{\alpha}$), and the direction of the vector $\vec{\omega}_{\alpha}$ corresponds to that of the normal vector $\vec{n}_{\alpha}$. Therefore, the following equation is obtained for the circulation of the $k$th vortex:
\begin{eqnarray}
 \Gamma_{k} &\simeq& \sum^{N^{(k)}_{p}}_{\alpha = 1} 2\omega_{\alpha}\frac{m_{\alpha}}{\rho_{\alpha}},~~~(\omega_{\alpha}=|\vec{\omega}_{\alpha}|). \label{eq:quantumcirdiscstokesfin}
\end{eqnarray}

\subsection{Magnus effect in SPH formalism}
An object rotating in a flow receives a lift force in a direction perpendicular to the direction of the flow, called the Magnus force. The magnitude of the Magnus force $F_{mg}$ in two-dimensional space is expressed as follows, according to the Kutta--Joukowski theorem:
\begin{eqnarray}
F_{mg} &=& \rho_{f} V_{f} \Gamma, \label{eq:KuttaJoukowski}
\end{eqnarray}
where $\rho_{f}$ is the fluid density, $V_{f}$ is the speed of the flow around the object, and $\Gamma$ represents the circulation.
In our case, as each vortex is composed of normal fluid particles, the velocity of the flow around a vortex can be obtained by sampling the velocities of the superfluid particles in the neighborhood of the normal fluid particles in the vortex. By using the circulation in Eq.~(\ref{eq:quantumcirdiscstokesfin}), the magnitude of the Magnus force acting on the $k$th vortex $F^{(k)}_{mg}$ is expressed as follows:
\begin{eqnarray}
F^{(k)}_{mg} &=& \rho_{s} \Biggl|\Biggl| \frac{1}{N^{(k)}_{p}} \sum^{N^{(k)}_{p}}_{\alpha = 1}\frac{1}{N^{(k, \alpha)}_{p}}\sum^{N^{(k, \alpha)}_{p}}_{\beta \in \Omega_{\alpha}} \vec{v}^{(\beta)}_{s} \Biggr|\Biggr|~{\rm min}~[\Gamma_{k}, ~q\kappa], \label{eq:KuttaJoukowskiSPH}
\end{eqnarray}
where $\rho_{s}$ is the mass density of the superfluid components. $N^{(k)}_{p}$ is the number of particles constituting the $k$th vortex. $N^{(k, \alpha)}_{p}$ is the number of superfluid particles in the neighborhood of the $\alpha$th particle inside the $k$th vortex.  $\vec{v}^{(\beta)}_{s}$ is the velocity of the $\beta$th particle among  $N^{(k, \alpha)}_{p}$ superfluid particles. $\rm min~[\cdots]$ is \HLLL{introduced} to avoid overestimating the Magnus force of each vortex that maintains the circulation $q\kappa$. 
The orientation of $F^{(k)}_{mg}$ is set by the vector $\vec{n}_{k} \times \vec{V}_{k}$, where $\vec{V}_{k}$ is the averaged velocity obtained via the calculation of the absolute value symbol in Eq.~(\ref{eq:KuttaJoukowskiSPH}), and $\vec{n}_{k}$ is a vector oriented in the vertical direction to the paper surface in two-dimensional simulations. 

\subsection{Modeling of the interaction forces among vortices}
The interaction energy per unit length $E^{(k,l)}_{int}$ between the $k$th and $l$th parallel vortices separated by a distance $d$ can be estimated as follows~\cite{pethick2008bose}: 
\begin{eqnarray}
E^{(k,l)}_{int} &\simeq& \frac{2\pi q_{k}q_{l} \hbar^2 n}{m}{\rm ln}\frac{D}{d},~~~(D \gg d,~d \gg \xi_{h}). \label{eq:vvintenergy}
\end{eqnarray}
Here, $\hbar$ represents the reduced Planck's constant. $q_{k}$ and $q_{l}$ are the winding numbers of the $k$th and $l$th vortices, respectively. $m$ is the mass of a helium-4 atom. $n$ is the density of the condensate. $D$ is a finite large distance; in the case of rotating liquid helium-4, the outer radius of the cylindrical vessel is set to $D$. $d$ is defined as the distance from the $k$th vortex. Note that Eq.~(\ref{eq:vvintenergy}) holds when $D \gg d$ and $d \gg \xi_{h}$, where $\xi_{h}$ represents the healing length of a vortex.
The interaction force that acts on the $k$th vortex from the $l$th vortex can be given by the first partial derivative of the interaction energy $E^{(k,l)}_{int}$ with respect to $d$ as follows:
\begin{eqnarray}
f^{(k,l)}_{int}&=&\frac{\partial E^{(k,l)}_{int}}{\partial d} \nonumber ~~\simeq~~ -\frac{1}{d} \frac{m}{2\pi} \cdot \Biggl(\frac{2\pi q_{k}\hbar}{m}\Biggr) \cdot \Biggl(\frac{2\pi q_{l}\hbar}{m}\Biggr) \cdot n \nonumber \\
	   	       &=& -\frac{1}{d} \frac{m}{2\pi} \cdot q_{k}\kappa \cdot q_{l}\kappa \cdot n. \label{eq:vvintforcebreakdown}
\end{eqnarray}
Here, we used the relation $\kappa = 2\pi\hbar/m$. The negative sign in Eq.~(\ref{eq:vvintforcebreakdown}) indicates that $f^{(k,l)}_{int}$ is a repulsive force. 
$q_{k}\kappa$ and $q_{l}\kappa$ provide the circulations of the $k$th and $l$th vortices, respectively. 
$n$ represents the density of the condensate, and $mn$ corresponds to the mass density $\rho_{s}$. Therefore, Eq.~(\ref{eq:vvintforcebreakdown}) can be expressed as
\begin{eqnarray}
f^{(k,l)}_{int} &\simeq& -\frac{1}{d} \Gamma_{k} \Gamma_{l} \frac{\rho_{s}}{2\pi}, \label{eq:vvintforcefin}
\end{eqnarray}
where the circulations $\Gamma_{k}$ and $\Gamma_{l}$ are computed using Eq.~(\ref{eq:quantumcirdiscstokesfin}).

We further approximated Eq.~(\ref{eq:vvintforcefin}) by replacing the distance $d$ with $d-\xi_{h}$ to consider the volume effect of each vortex. This approximation is acceptable because Eq.~(\ref{eq:vvintenergy}) becomes valid when $d \gg \xi_{h}$, by which the relationship of $d \approx d - \xi_{h}$ is obtained. The resulting formula is as follows:.
\begin{eqnarray}
f^{(k,l)}_{int} &\approx& -\frac{1}{d-\xi_{h}} \Gamma_{k} \Gamma_{l} \frac{\rho_{s}}{2\pi}. \label{eq:vvintforcefinaprox}
\end{eqnarray}
Here, as the healing length $\xi_{h}$ provides the scale of a vortex, we \HL{determine} $\xi_{h}$ at the initial state from the average volume of a vortex and the average number of particles in a vortex on the premise of the circular shape and hold the initial value of $\xi_{h}$ during the simulation.

\subsection{Representation of the quantization of circulation}
In quantum hydrodynamics, the circulation of each vortex is given by the integral multiple of the quantum of circulation $\kappa$, as represented in Eq.~(\ref{eq:quantumcirkappa}). Under such topological defects, quantum lattices are formed owing to the balance between the repulsive forces among multiple parallel vortices and the Magnus forces, which operate as centripetal forces owing to the forced rotation. Therefore, modeling the quantization of circulation in SPH formalism is critical for reproducing quantum lattices in simulations. 
This study introduces the following conditional judgment as to $f^{(k,l)}_{int}$ to simulate the effect of quantization of circulation:
\begin{eqnarray}
 F^{(k,l)}_{int} &=& 
~\Biggl\{ \begin{array}{l}
{0~~~~~~~~(\Gamma_{k} < q \kappa,~~~\Gamma_{l} < q \kappa),} \\
{f^{(k,l)}_{int}~~~{\rm (otherwise),}} \\
	  \end{array}
	\label{eq:vvforcejudgement}
\end{eqnarray}
where $ F^{(k, l)}_{int}$ is the resulting interaction force between the $k$th and $l$th vortices obtained after the judgment. 
Equation~(\ref{eq:vvforcejudgement}) suggests that a bundle of normal fluid particles is identified as a vortex when its circulation becomes greater than $q \kappa$, whereas it is recognized as a cluster otherwise. In the simulations, the repulsive force $f^{(k,l)}_{int}$ is computed between the $k$th vortex and the $l$th vortex or cluster to prevent the $k$th vortex from further development. In contrast, the interaction forces between clusters with circulations less than $q \kappa$ are ignored because they do not have sufficient influence on each other. Finally, the sum of the interacting forces acting on the $k$th vortex $F^{(k)}_{int}$ is expressed as
\begin{eqnarray}
F^{(k)}_{int} = \sum^{N_{v}}_{l}  F^{(k,l)}_{int}, \label{eq:intforcefin}
\end{eqnarray}
where $N_{v}$ indicates the number of vortices detected at each computational step. The Magnus forces and the interaction forces among the vortices are then added to the motion equations of Eq.~(\ref{eq:goveqsuper:mut}) and Eq.~(\ref{eq:goveqnormal:mut}) as external forces. Specifically, the acceleration obtained by dividing $F^{(k)}_{mg}$ and $F^{(k)}_{int}$ by the vortex mass is evenly distributed to each particle belonging to the vortex. The system is then updated using an explicit time-integrating scheme, and all the processes for a computational time step are completed.

\begin{figure*}[t]
\vspace{-14.0cm}
\hspace{8.5cm}
\centerline{\includegraphics[width=2.1\textwidth, clip, bb= 0 0 1536 1125]{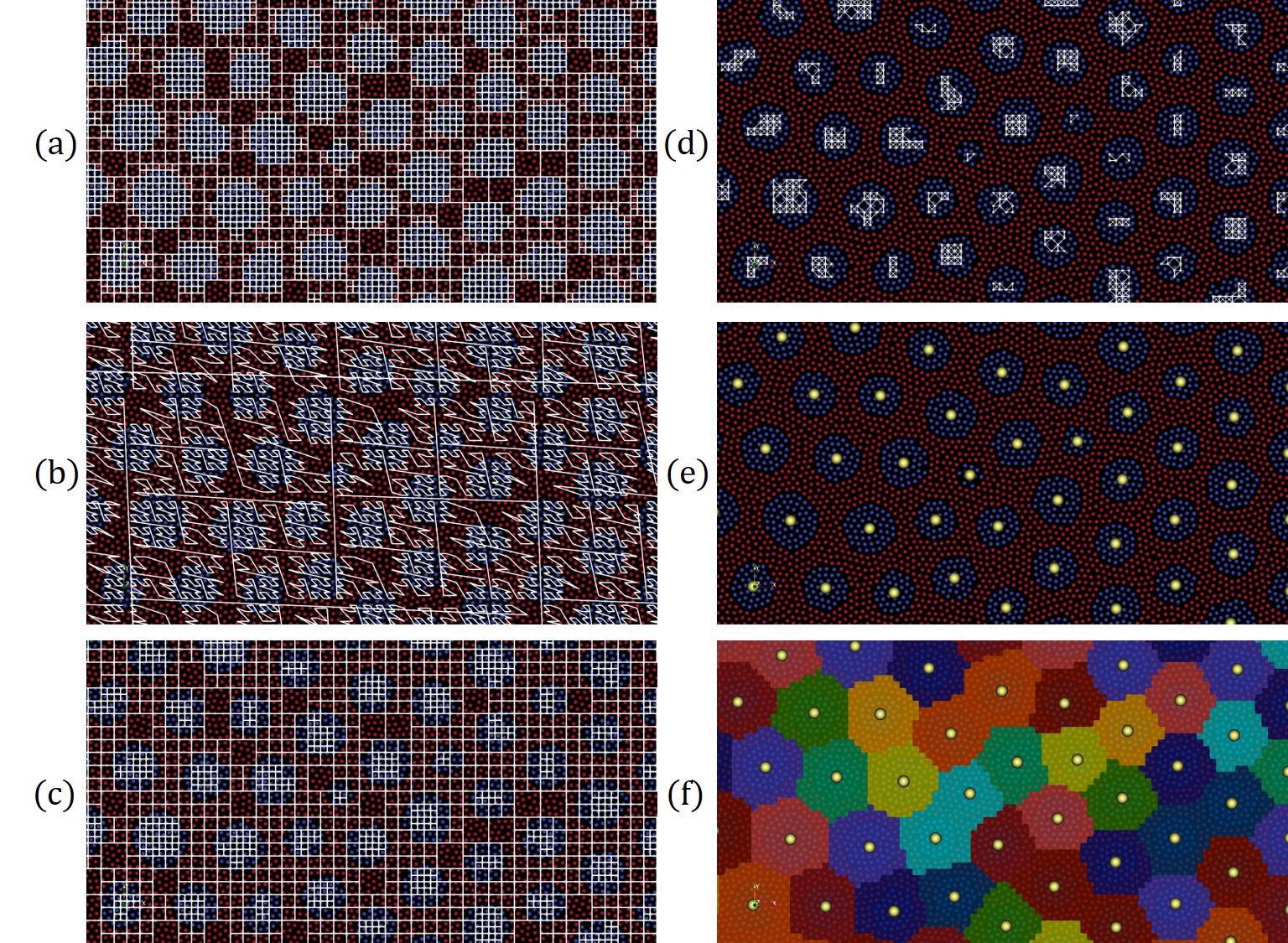}}
\caption{Schematic of our vortex detection method: \HLLL{(a) an example of the generated refined meshes, (b) the process of tracing the distal leaf using a z-curve, (c) the meshes modified using the thinning technique, (d) the bundles of leaves for each vortex, (e) the coordinates of the real center points of the vortices indicated by the yellow circles, and (f) the result of Voronoi partitioning}}
\label{fig:Figure_VoronoiDetection}
\end{figure*}

\section{Key computational algorithms for practical simulations}
In actual simulations, it is indispensable to identify clusters composed of normal fluid particles as individual groups based on their geometric arrangements and set sequential numbers at each simulation step. A suitable partitioning algorithm for classifying particles into groups is required. The following section presents a vortex detection algorithm using Voronoi tessellation and a hierarchically structured quadtree.

\subsection{Vortex detection algorithm using Voronoi tessellation and a hierarchically structured quadtree}
Our vortex detection algorithm is implemented using the following five computational procedures: (a) the recursive mesh refinement of the computational domain, (b) the topological thinning of the quadtree, (c) the connection of adjacent deepest distal leaves, (d) the estimation of the center points of vortices, and (e) Voronoi tessellation. 
The details of each procedure are presented in Fig.~\ref{fig:Figure_VoronoiDetection}. The subfigures in Fig.~\ref{fig:Figure_VoronoiDetection} show the same particle distribution with superfluid particles in red and normal fluid particles in blue on the back.

\paragraph{(a) Recursive mesh refinement} 
The computational domain is recursively divided into $2\times2$ square-shaped subleaves using a quadtree until the number of normal fluid particles in each leaf becomes less than or equal to one. Figure~\ref{fig:Figure_VoronoiDetection}a shows an example of the generated refined meshes with a white-colored line over the aforementioned particle distribution.

\paragraph{(b) Topological thinning of the quadtree} 
All the distal leaves of a hierarchically structured quadtree generated by process (a) are traced by the Morton curve (z-curve), a space-filling curve that fills in a space with a single stroke of a brush~\cite{konrad:inria-00331382, SFCDecomp1997-Aluru, Tsuzuki:2016:EDL:3019094.3019095, 4658143}. Figure~\ref{fig:Figure_VoronoiDetection}b shows the process of tracing the distal leaf using a z-curve. Next, the deepest distal leaves of the tree, which are adjacent to the shallower leaves covering the region of the superfluid components, are released from the outer leaves while tracing all the distal leaves using the z-curve. This method is known as a topological thinning technique in the field of image processing. Please refer to the literature~\cite{TSUZUKI2019161} for further details on the process of releasing leaves. The use of the thinning technique facilitates distinguishing different vortices, even when they are close to each other. We set the minimum size of the refined meshes to the initial particle distance $d_{p}$. Therefore, the vortex can be distinguished with a resolution of $2d$ because the two leaves are released in each direction. Figure~\ref{fig:Figure_VoronoiDetection}c shows the meshes modified using the thinning technique. The outer leaves in each bundle of distal leaves were confirmed to be released, as shown in Fig.~\ref{fig:Figure_VoronoiDetection}a.

\paragraph{(c) Connection of adjacent deepest distal leaves} Each deepest distal leaf searches for other deepest distal leaves located within a distance of ${\rm 1.5}d_{p}$ from itself and registers them in its list. The value of ${\rm 1.5}d_{p}$ was determined to detect the closest leaf in the diagonal direction, in addition to the vertical and horizontal directions. We then searched for the connections among the leaves using Algorithm~\ref{alg:searchconnect}.
\begin{algorithm}[h]
\caption{Search for the connections among the neighboring deepest distal leaves} \label{alg:searchconnect}
\begin{algorithmic}[1]
{\small
\State {${\tt k} \leftarrow 0$}  
\For{${\tt j}$ = 0 $\to$ ${\tt N_{leaf}}$}
\If{{\tt Leaf[j].reg}~$=$~{\tt false}} 
\State {Add~{\tt Leaf[j]}~to the list of ~{\tt Vortex[k]}} 
\State {{\tt Leaf[j].reg}~$\leftarrow$~{\tt true}}
\State {Check the status of each {\tt Leaf[$\cdots$]} on the list of {\tt Leaf[j]} and add it to the list of~{\tt Vortex[k]} if unregistered.} 
\State {/* Execute line 6 recursively until we can no longer find an unregistered leaf on the chain. */}
\State {${\tt k} \leftarrow {\tt k+1} $}  
\EndIf
\EndFor
}
\end{algorithmic}
\end{algorithm}

In Algorithm~\ref{alg:searchconnect}, $N_{leaf}$ represents the number of deepest distal leaves (hereafter called ``leaves''). The parameter $k$ represents the index of vortices. 
We start with the case of $k=0$. Initially, each flag {\tt reg} of all leaves is set to false because no leaf is registered to any vortex. First, we register the $0$th leaf to the $0$th vortex and change the flag {\tt reg} of the $0$th leaf to true, as in line 5. Next, we recursively find all the leaves that have direct or indirect paths to the $0$th leaf, and we add them to the list of the $0$th vortex similarly. A more specific procedure is described as follows: As the $0$th leaf holds a list of the leaves located within a distance of $1.5d_{p}$, it checks whether each leaf on its list is already on the list of the $0$th vortex. If it finds an unregistered leaf, it adds the found leaf to the list of the $0$th vortex. The newly added leaf changes the flag {\tt reg} to true, starts the same search, and recursively registers leaves to the $0$th vortex. By this chain access, all the leaves that are directly or indirectly connected to the $0$th leaf are added to the list of the $0$th vortex. All the procedures are continued on the chain until unregistered leaves can no longer be found on the $0$th vortex. Thereafter, we repeat the same process for the $1$st vortex. At this time, the leaves that are already registered to the $0$th vortex are left out from the candidates for the $1$st vortex by the conditional judgment in line 3. After all the aforementioned processes are completed, the value of $ k $ determines the number of vortices $N_{v}$. We can confirm from Fig.~\ref{fig:Figure_VoronoiDetection}d that there is a bundle of leaves for each vortex, and the leaves of each bundle are connected to each other in the horizontal, vertical, and diagonal directions.

\paragraph{(d) Estimation of the center points of vortices}
We calculate the temporary center points of the vortices using the arithmetic mean of the center coordinates of the deepest distal leaves in each bundle. These center points are calculated only to create the Voronoi diagram in the following procedure (e), and the real center points of the vortices are calculated after completing procedure (e) before computing the interactions among the vortices. Figure~\ref{fig:Figure_VoronoiDetection}e shows the coordinates of the real center points of the vortices indicated by the yellow circles. Notably, the possible deviation between a temporary center point and the corresponding real center point becomes approximately $d_{p}$ because the minimum size of the refined meshes is set to $d_{p}$ in this method. 

\paragraph{(e) Voronoi tessellation}
Using the temporary center point of each vortex as a generating point of the Voronoi diagram, we perform Voronoi tessellation for the simulation domain covered by a background mesh whose cell size is given by $d_{p}$. Let us define a set of generating points as $P=\{p(1),p(2),p(3)\cdots,p(N_ {v})\}$, and denote the $k$th Voronoi area as $V(k)$, which satisfies $V(k) = \{ q \in Q \mid d[p(k), q] \le d[p(l), q], p(l) \in P\}$, where $Q$ is the set of cells in the background mesh; and $d[x,y]$ represents the Euclidean distance between $x$ and $y$. If the generating point $p(k)$ is the closest to the cell $q$ in the background mesh, we can conclude that the cell $q$ belongs to $V(k)$ and thus store the index $k$ to the cell $q$ of the background mesh. 
Figure~\ref{fig:Figure_VoronoiDetection}f shows the result of performing the Voronoi partitioning using the method described above. From the comparison with the particle distributions on the back, it can be confirmed that each vortex is included in each Voronoi area. Note that the same color occasionally shows different areas because only 12 colors are used. In this method, when a large vortex and a small vortex are adjacent to each other, the large vortex may partially protrude from the Voronoi area as an exceptional case. However, it is unlikely that only a specific vortex becomes large in practical simulations because the repulsive forces act between the vortices once the circulation exceeds $qk$, as expressed in Eq.~(\ref{eq:vvforcejudgement}). 
	
After the computational procedure from (a) to (e), each vortex calculates the sum of the circulations from its particles using Eq.~(\ref{eq:quantumcirdiscstokesfin}) and computes the Magnus force acting on the vortex using Eq.~(\ref{eq:KuttaJoukowskiSPH}). We then compute the repulsive forces among the vortices using Eq.~(\ref{eq:intforcefin}) using the circulations and real center points. After each vortex distributes the forces to its particles, we update the velocities and positions of each particle using the velocity Verlet method, as previously mentioned.

\section{Numerical simulations of rotating liquid helium-4}
This section deals with the problem of rotating liquid helium-4 under the same numerical conditions as those reported in the literature~\cite{Tsuzuki_2021}. Here, we perform the simulations of rotating liquid helium-4 to determine whether to reproduce the phenomena of vortex lattices in addition to the rigid-body rotation \HLLL{and the spinnings of the vortices}, in both the current and previous models.

\subsection{Setting of parameters} 
Let us set the outer diameter of a cylindrical vessel to $\rm 0.2~cm$ and the rotational angular velocity around the cylinder axis to $\rm 5~rad \cdot s^{-1}$ counterclockwise. The resolution of $(N_{px}, N_{py})$ is given by $\rm (500, 500)$, where $N_{px}$ and $N_{py}$ represent the number of particles in the $x$ and $y$ directions, respectively, in a square space that covers the cylindrical container. The temperature $T$ is set to $\rm 1.6~K$, and the parameters $(\rho_{s}/\rho, \rho_{n}/\rho, \kappa)$ are determined by referring to the values at $\rm T = 1.6~K$ listed in the literature~\cite{Vinen2002}. The resulting values are given by $\rm (0.834, 0.166, 0.001)$. In addition, $\eta_{n}$ is obtained from the relationship of $0.566~\kappa\rho_{n}$ with reference to~\cite{Vinen2002}. The value of $\eta_{n}$ is set as the shear viscosity $\bar{\eta}$ and rotational viscosity $\bar{\eta}_{r}$. The winding number $q$ is set to $\rm 1$. The parameters $(\alpha, \beta_{v}, L_{0})$ of a decay model for the vortex line density are determined with reference to~\cite{NEMIROVSKII201385}.  
In contrast, $C_{\omega}$ is an empirical parameter for controlling the magnitude of angular velocity $\vec{\omega}$. 
First, we introduce two cases, namely, $C_{\omega}={\rm 0.01}$ and $C_{\omega}={\rm 0.005}$, to investigate the influence of the parameter $C_{\omega}$ on the simulation results. 
The normal fluid particles were placed along the inner circumference of the container as walls. No-slip conditions were imposed on the wall particles as boundary conditions. This is \HLLL{appropriate} because both the normal and superfluid components exert frictional forces~\HLL{(viscous friction forces for the former and mutual friction forces for the latter)} against the normal fluid particles that form the walls. In the initial state, superfluid or normal fluid particles were randomly generated in proportion to the density ratio and distributed in the vessel.

\begin{figure*}[t]
\vspace{-6.5cm}
\hspace{9.0cm}
\centerline{\includegraphics[width=2.1\textwidth, clip, bb= 0 0 2000 684]{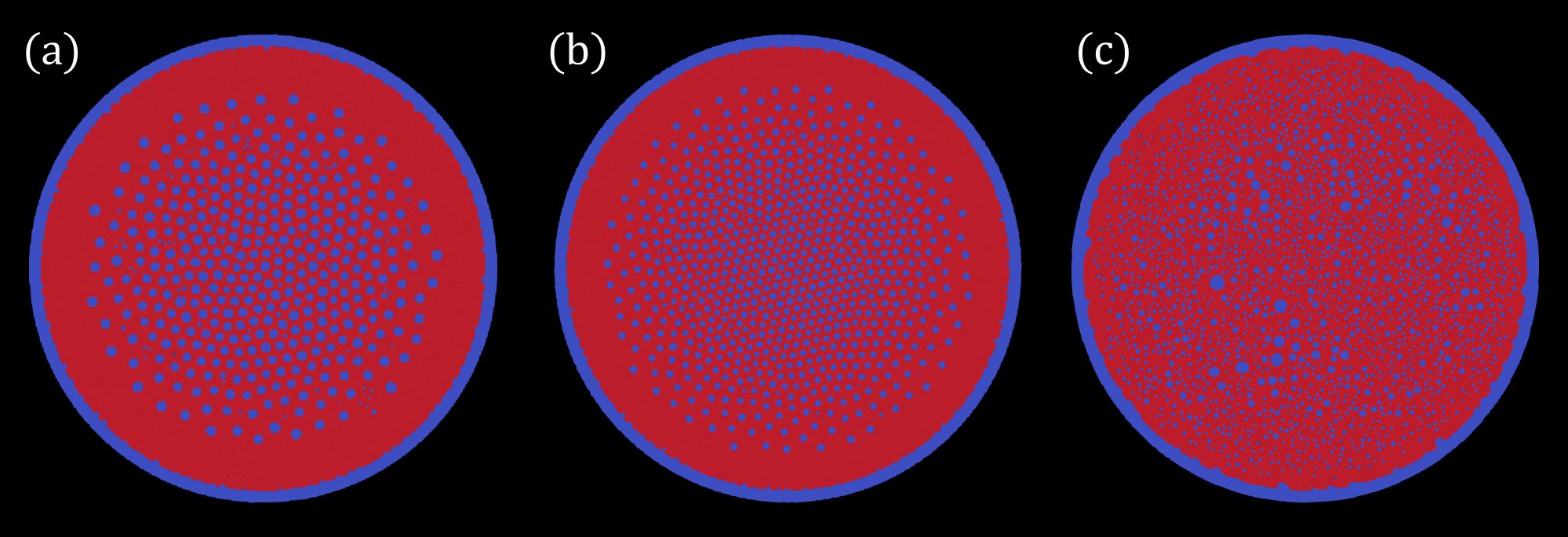}}
\caption{Snapshots of simulations after \HL{convergence is reached}. Each subfigure shows the simulation results (a) when $C_{\omega}$~=~${\rm 0.005}$, (b) when $C_{\omega}$~=~${\rm 0.01}$, and (c) when $C_{\omega}$~=~${\rm 0.01}$ using our previous model for comparison.}
\label{fig:Figure_RotatingLiquid4He}
\end{figure*}

\subsection{Simulation results} 
\begin{figure}[t]
\vspace{-8.0cm}
\hspace{8.0cm}
\centerline{\includegraphics[width=1.9\textwidth, clip, bb= 0 0 2889 1391]{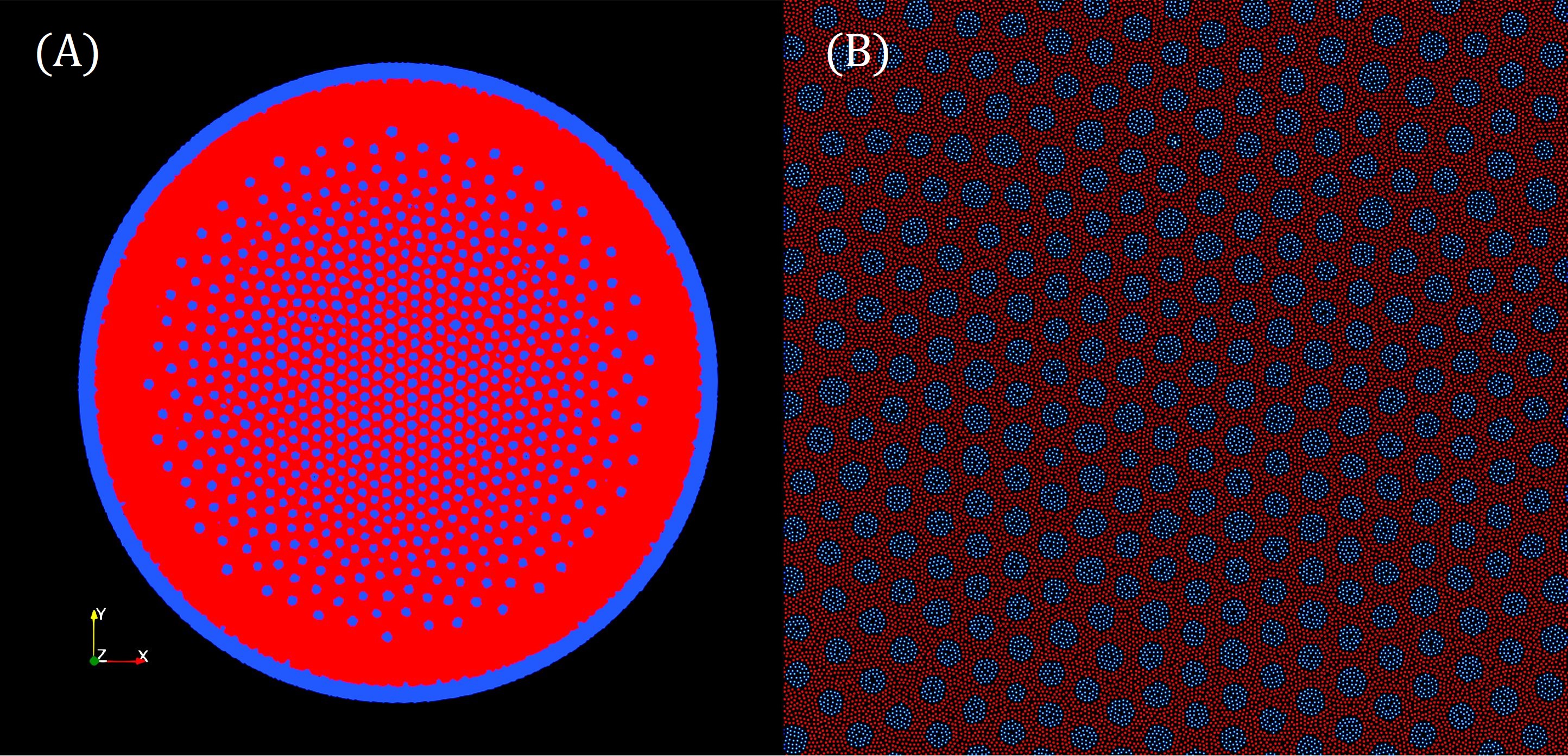}}
\caption{\HLLL{A video of (A) the arrangement in the case of Fig.~\ref{fig:Figure_RotatingLiquid4He}b and (B) its enlarged view~(Multimedia view)}}
\label{fig:Figure_VideoImm}
\end{figure}

Figure~\ref{fig:Figure_RotatingLiquid4He} shows the snapshots of the simulations after \HL{9} s in physical time, \HLLL{which consumed approximately 67 hours in computational time on a GPU NVIDIA Geforce RTX 2080 Ti.} Each subfigure shows the simulation results (a) when $C_{\omega}$~=~${\rm 0.005}$, (b) when $C_{\omega}$~=~${\rm 0.01}$, and (c) when $C_{\omega}$~=~${\rm 0.01}$ using our previous model for a comparison. After a short time elapsed from the start of the simulation, the normal fluid particles, which are low-density components at this temperature range, gathered to form clusters and began rotating in a direction parallel to the rotation axis of the outer cylindrical vessel. The processes up to this point were the same among (a), (b), and (c). However, in the cases of (a) and (b), it was observed that the generated clusters moved toward the centripetal direction upon receiving the Magnus forces generated by their own rotations, whereas they began to repel each other once they became vortices as their respective circulations reached $\kappa$. Owing to the balance between the Magnus force and the repulsive forces among the vortices, the characteristic regular arrangements of vortices were observed to form as shown in (a) and (b). A video of \HL{the arrangement} in the case of (b) \HLLL{and its enlarged view} is provided as \HLLL{integral multimedia in Fig.~\ref{fig:Figure_VideoImm}}.
\HLLL{It was also confirmed from the video that several small clusters of normal fluid particles which adhered to the walls in the initial state and remained on them started to exfoliate from the wall because of the centripetal forces caused by the forced rotation of the outer cylindrical vessel.}

\begin{figure*}[t]
\vspace{-5.2cm}
\hspace{9.0cm}
\centerline{\includegraphics[width=2.1\textwidth, clip, bb= 0 0 2002 636]{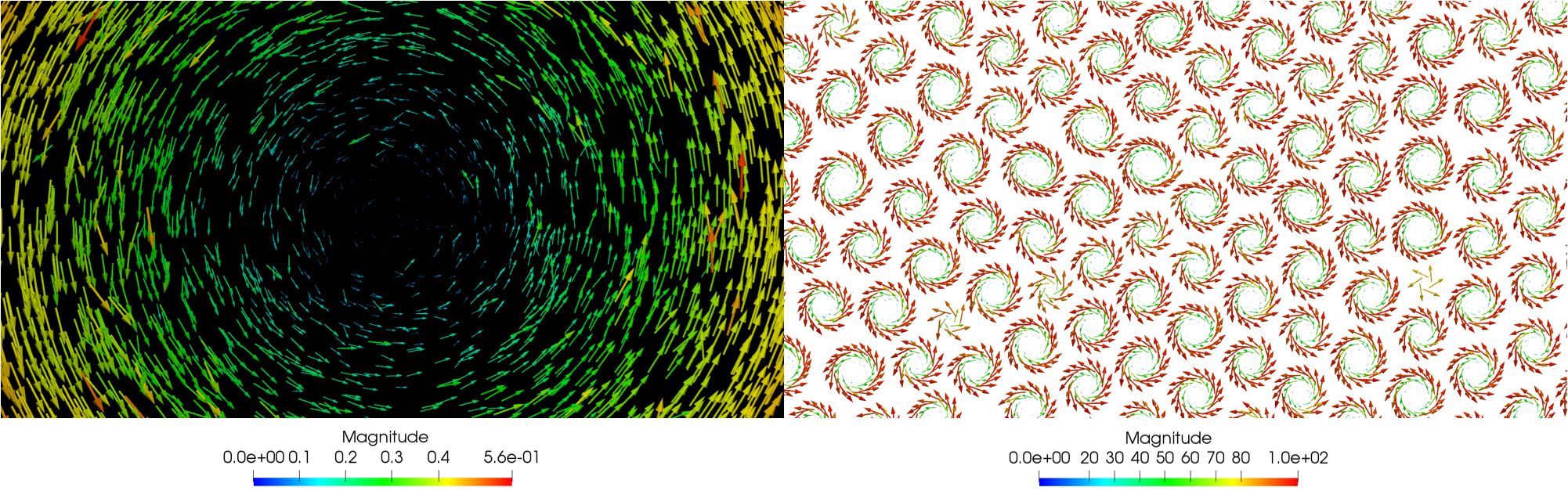}}
\caption{Example snapshot of the rotational force generated on each vortex for $C_{\omega}$~=~$\rm 0.01$~(right part) and the entire velocity field at that time~(left part), which suggests that the entire velocity field is irrelevant to the tiny spinning of vortices on the overall scale. }
\label{fig:Figure_RotSimVelocity}
\end{figure*}

In addition, the spinnings of the vortices were \HL{successfully} observed in the respective cases. The right side of Fig.~\ref{fig:Figure_RotSimVelocity} shows an example snapshot of the rotational force generated on each vortex in the case of (b). The length and color of each arrow are proportional to the magnitude of the rotational force. The left side of Fig.~\ref{fig:Figure_RotSimVelocity} represents the velocity field at the same time as the right side of the figure on the overall scale, where the length and color of each arrow are proportional to the magnitude of the velocity. This suggests that the velocity field is approximately irrelevant to the tiny spinnings of the vortices on the overall scale. \HL{The} spinnings of vortices were observed even in the case of $C_{\omega}$ $<$ ${\rm 0.005}$ as in (a) and (b), whereas they were scarcely observed in the case of (c) because of the imbalance among the forces due to the lack of vortex dynamics. 

\begin{figure*}[t]
\vspace{-8.0cm}
\hspace{7.5cm}
\centerline{\includegraphics[width=1.8\textwidth, clip, bb= 5 5 1170 592]{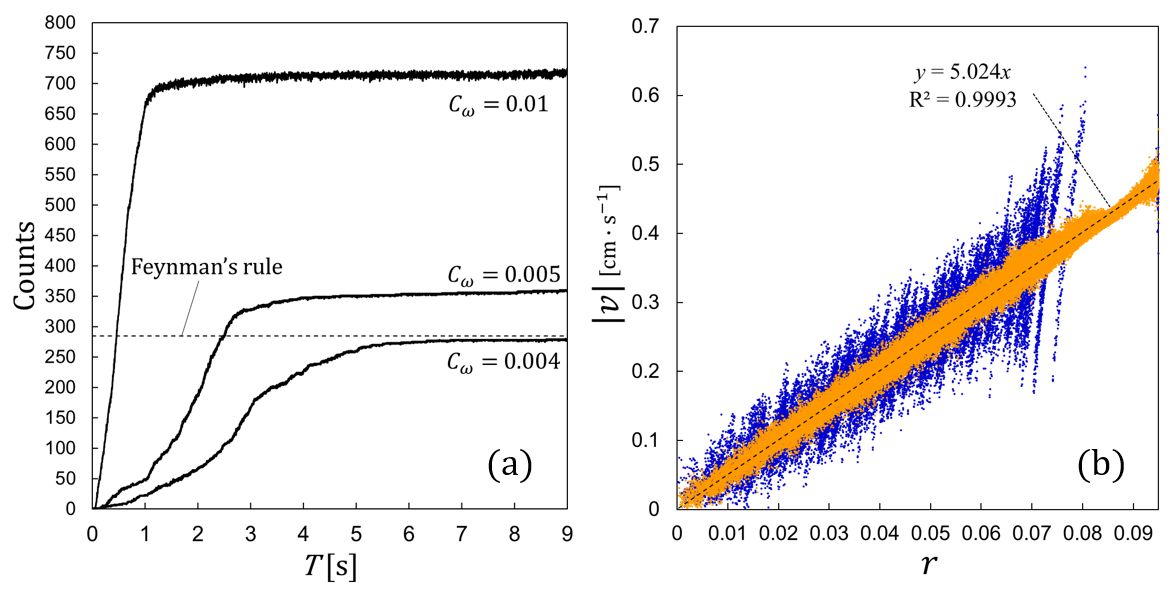}}
\caption{Breakdown of the simulation results: (a) Time dependence of the number of vortices in the three cases of $C_{\omega}={\rm 0.01}$, $C_{\omega}={\rm 0.005}$, and $C_{\omega}={0.004}$; (b) Velocity distribution of normal fluid particles (blue) and that of superfluid particles (yellow) when $C_{\omega}={0.004}$
}
\label{fig:Figure_ResultsA}
\end{figure*}

We denote the number of vortices per unit area generated in the rotating liquid helium-4 in a cylindrical container as $n_{v}$. According to Feynman's rule~\HLLL{\cite{feynman1955progress, VANSCIVER2009247}}, $n_{v}$ follows the relationship $n_{v} = 2\Omega/\kappa$, where $\Omega$ is the angular velocity of the outer cylindrical vessel, and $\kappa$ is the quantum of circulation. Notably, $\Omega$ and $\kappa $ were set to $\rm 5~rad\cdot s^{-1}$ and $1.0\times10^{-3} {\rm cm^{2}\cdot s^{-1}}$, respectively, which yields $n_{v}$ = $1.0\times 10^{4}~{\rm cm^{-2}}$. The ideal number of vortices estimated by multiplying the inner cylinder area by $n_{v}$ is approximately $\rm 284.7$. Meanwhile, image analysis using {\tt ImageJ}~\cite{10030139275, abramoff2004image, Schneider2012} revealed the number of vortices in each case. Figure~\ref{fig:Figure_ResultsA}a shows the dependence of the number of clusters whose circulations are greater than $\kappa$, that is, the number of detected vortices, on the elapsed time. \HL{Interestingly}, the number of generated vortices increased as $C_{\omega}$ increased. We first performed two cases of $C_{\omega}={\rm 0.005}$ and $C_{\omega}={\rm 0.01}$ and then estimated the value of $C_{\omega}$, where the number of vortices corresponds to Feynman's rule by extrapolating the linear relationship between the two cases at the plateau to be $\rm 0.004$. Figure~\ref{fig:Figure_ResultsA}(a) shows the time dependence of the number of vortices when $C_{\omega} = {\rm 0.004}$. It can be confirmed from Fig.~\ref{fig:Figure_ResultsA}a that the number of vortices is close to the theoretical value when $C_{\omega} = {\rm 0.004}$; the parameter $C_{\omega}$ functions as a control parameter for the number of vortices in the system. 

\begin{figure}[t]
\vspace{-8.0cm}
\hspace{7.5cm}
\centerline{\includegraphics[width=1.8\textwidth, clip, bb= 10 10 1193 618]{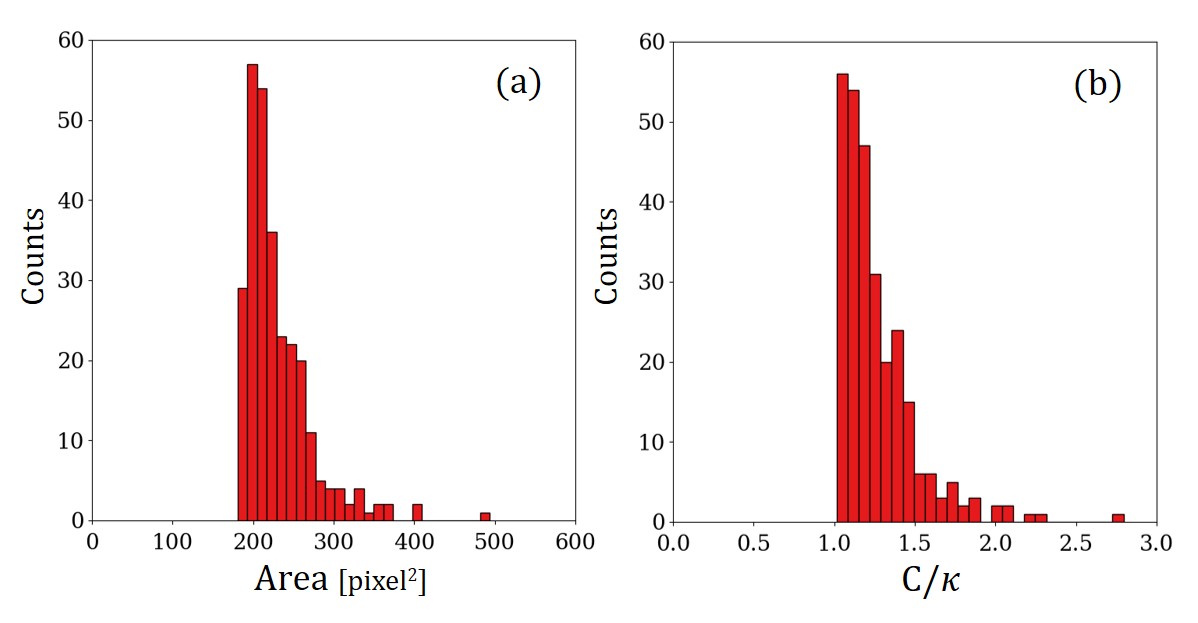}}
\caption{Histograms of \HLLL{(a)} the areas of vortices and \HLLL{(b)} the $\kappa$-scaled circulations of vortices, for $C_{\omega}={0.004}$}
\label{fig:Figure_ResultsB}
\end{figure}

Figure~\ref{fig:Figure_ResultsA}(b) shows the dependence of the velocity distribution of superfluid (yellow) and normal fluid particles (blue) on the distance from the center of the cylindrical vessel. The velocities of the normal fluid particles are confirmed to fluctuate compared with those of the superfluid particles due to the spinnings of vortices. \HL{However, as previously mentioned, they mainly contribute to the velocity in the spinning direction and hardly to the entire velocity field}. The dashed line represents the line fitted to the velocity distribution obtained using the least square method, which suggests that the averaged velocity 
approximately corresponds to the theoretical case $y = 5.0x$ obtained from the relationship of $|v|=r|\Omega|$ when setting the angular velocity of the outer vessel $\Omega$ to $\rm 5~rad \cdot s^{-1}$; the system undergoes rigid-body rotation.
Figure~\ref{fig:Figure_ResultsB}(a) shows a histogram of the areas of the vortices when $C_{\omega}$ = ${\rm 0.004}$ obtained using {\tt ImageJ}~\cite{10030139275, abramoff2004image, Schneider2012}. pixel$^2$ represents the unit of area.
The histogram should tend to a delta function because the vortices become uniform in the ideal case. However, the width of the histogram was not sufficiently \HLLL{narrow}. A possible reason for this discrepancy is that the circulation of the vortices is not well quantized \HL{even in the improved model}. Figure~\ref{fig:Figure_ResultsB}(b) shows the histogram of the circulation of the vortices scaled by $\kappa$ when $C_{\omega}$ = ${\rm 0.004}$. It was observed that the distribution range changed continuously, unlike in \HLLL{Vinen's} \HL{experiment}~\cite{vinen1961detection}. 

\HLLL{Furthermore, the following observations were made: First, the mode value of $C/\kappa$ was approximately $\rm 1.048$ and $\rm 1.02$ in the simulation and experiment~\cite{vinen1961detection}, respectively; therefore, the percentage deviation between each mode value of the simulation and experiment and the ideal value of 1 can be estimated as approximately 4.8 and 2 percent, respectively. Here, the theoretical value of $C/\kappa$ becomes an integer, and it most frequently becomes one~\cite{pethick2008bose}. Second, the experiment had a nearly symmetric histogram to the mode value, while the simulation had a histogram without the tail area on the left side of the mode, similar to a half-normal distribution; in other words, the skewness tends to be approximately zero in the experiment and positive in the simulation. Moreover, it was observed that approximately 2.5 percent of the total vortices had circulations of $C/\kappa \ge 2$ in the simulation, which was not observed in the experiment.}
\HLLL{To summarize}, the observed vortex lattices cannot be regarded as ``quantum lattices'' at this stage of the research; a stronger constraint is required in addition to the conditional judgment by Eq.~(\ref{eq:vvforcejudgement}) to represent the circulation quantization more accurately. 

\HL{Consequently}, we \HL{qualitatively} reproduced several phenomena of rotating liquid helium-4~(the emergence of multiple parallel vortices and their rigid-body rotations, and the form of vortex lattices) by solving the \HL{reformulated} two-fluid model using SPH based on the continuum mechanics approximation. Further sophisticated models are necessary for a more precise analysis. In particular, the accurate modeling of the quantization of circulation should be addressed first.
Further, other microscopic parameters require optimization, which should be conducted \HLLL{ based on the theories of many-body interactions and external fields~\cite{PhysRevLett.94.050402, PhysRevLett.101.010402, PhysRevA.81.025604}, the two-fluid model of the normal fluid and Bose-Einstein condensates with microscopic quantum interactions~\cite{doi:10.1063/5.0053035},} or experimental studies\HLLL{~\cite{vinen1961detection, doi:10.1063/1.4991558, doi:10.1063/1.4984913, doi:10.1063/1.4940980}}. Although our current model is still a ``toy model,'' it provides a good perspective for the feasibility of approaching these interesting phenomena from a different perspective.

\section{Conclusions}
Our recent study has reported that the representative phenomena of liquid helium-4 rotating in a cylinder could be simulated by solving the two-fluid model using SPH after reformulating the viscosity to conserve the rotational angular momentum. Specifically, the emergence of multiple parallel vortices and their rigid-body rotations were observed in our previous SPH simulations. The reported model is based on a classical approximation that assumes fluid forces for both components and their interactions, with the expectation of functioning as a coarse-grained model of existing approximations that simultaneously solve the two components using a microscopic model~(e.g., vortex filament model) and the Navier--Stokes equation. 
Based on previous studies, this paper proposes an improved SPH scheme that explicitly incorporates vortex dynamics into the SPH formulation to reproduce \HL{vortex} lattices, which was not possible in previous studies. Consequently, our improved model was observed to reproduce vortex lattices by introducing the Magnus force and the interaction forces among vortices into the reformulated two-fluid model. 
The spinnings of vortices and their rigid-body rotations were also observed in the SPH simulations.
The number of vortices showed a certain agreement with Feynman's rule after the model parameter was optimized. 
Notably, from a scientific perspective, such vortex lattices are reproduced by a classical-mechanical approximation.

\HLLL{
From an application point of view, these findings are significant steps towards realizing a direct numerical simulation of bulk quantum liquids. As stated, liquid helium-4 has played a significant role in many scientific applications, as exemplified by the cryogenic cooling systems of astronomy satellites or space telescopes. Real-scale simulations of the superfluid or near-superfluid liquid helium inside a large-scale cryogenic cooling system can enable the streamlining of the development processes of these systems and realize their safer and more secure operation. However, presently, it is still unrealistic to directly solve microscopic relationships of all atoms constituting the bulk fluid. Notably, we successfully developed a numerical scheme based on continuum mechanical approximation that can reproduce the phenomena that were previously believed to be reproduced only through quantum-mechanical approaches by incorporating the vortex dynamics of a quantum fluid into the SPH formulation.} We hope that our model will help physicists studying low-temperature physics find a new way of approaching this bizarre phenomenon that has attracted attention for more than 80 years.

\section*{Acknowledgment}
This study was supported by a ``Grant-in-Aid for JSPS Fellows'' in Japan. The author would like to thank Editage (www.editage.jp) for English language editing.
The author would also like to express his gratitude to his family for their moral support and warm encouragement.

\bibliographystyle{h-physrev3}
\bibliography{reference}

\end{document}